\documentclass[aps,pre,showpacs,twocolumn,superscriptaddress,unsortedaddress]{revtex4}
\usepackage{bm}
\usepackage{graphicx}
\usepackage{amsmath}
\bibstyle{approve.bib}
\begin{document}
\title{A novel formulation of nonlocal electrostatics}
\author{A. Hildebrandt}
\email{anhi@bioinf.uni-sb.de}
\affiliation{Center for Bioinformatics, Saarland University, P.O. 15 11 50,
66041 Saarbr\"ucken, Germany}
\author{R. Blossey}
\affiliation{Interdisciplinary Research Institute c/o IEMN, Cit\'e
Scientifique BP 69, F-59652 Villeneuve d'Ascq, France}
\author{S. Rjasanow}
\affiliation{Dept. of Mathematics, Saarland University, P.O. 15 11 50, 66041 Saarbr\"ucken, Germany}
\author{O. Kohlbacher}
\affiliation{Dept. for Simulation of Biological Systems,    
   WSI/ZBIT, University of T\"ubingen, Sand 14, 72070 T\"ubingen, Germany}
\author{H.-P. Lenhof}
\affiliation{Center for Bioinformatics, Saarland University, P.O. 15 11 50,
66041 Saarbr\"ucken, Germany}
\date{\today}

\begin{abstract}
The accurate modeling of the dielectric properties of water is crucial for
many applications in physics, computational chemistry and molecular biology.
This becomes possible in the framework of nonlocal electrostatics, for which 
we propose a novel formulation allowing for numerical solutions for the 
nontrivial molecular geometries arising in the applications mentioned before. 
Our approach is based on the introduction of a secondary field, $\psi$, which 
acts as the potential for the rotation free part of the dielectric 
displacement field ${\bf D}$. 
For many relevant models, the dielectric function of the medium can be 
expressed as the Green's function of a local differential operator. 
In this case, the resulting coupled Poisson (-Boltzmann) equations for 
$\psi$ and the electrostatic potential $\phi$ reduce to a system of coupled 
PDEs. The approach is illustrated by its application to simple
geometries.
\end{abstract}

%\pacs{89.75.Hc, 87.23.Ge, 64.60.Ak, 05.90.+m}

\maketitle

The theory of continuum electrostatics plays a major role in the
determination of solvation free energies of atoms, ions, and 
biomolecules \cite{honig95}. Recent progress in 
its applicability to biological systems has been impressive:
the electrostatic potentials of large biomolecules such as, e.g., 
microtubuli and ribosomes, can be determined \cite{baker01}.
Unfortunately, the standard continuum approach ultimately becomes inaccurate 
when used to determine electrostatic properties on atomic scales 
\cite{simonson01}, as it is featureless, i.e., the correlation 
between solvent arrangements and the geometrical structure of 
biomolecular assemblies is not taken into account. 
On the other hand, continuum electrostatics is still much more efficient
from a computational point of view than microscopic simulations based on, e.g.,
molecular dynamics (MD). 
Therefore, interest has risen recently in extensions of the theory of 
continuum electrostatics that allow to account for spatial variations of the 
dielectric behaviour of the solvent, in particular near boundaries 
\cite{hansen,honig95}. Part of the motivation for such approaches stems from 
the field of protein docking, where a realistic and efficient modelling 
of solvent properties is essential \cite{docking}.

Within the continuum theory of electrodynamics, spatial dispersion effects 
can be taken into account in an approach called `nonlocal electrostatics'
\cite{korn1,korn2,korn3,korn4}. It rests on the assumption of a linear 
relationship between the dielectric displacement field and the electric 
field mediated by a permittivity kernel depending on two spatial arguments,
\begin{equation}\label{nonlocal}
{\bf D}({\bf r}) =
\varepsilon_0\int d{\bf r'} \varepsilon({\bf r},{\bf r'})
{\bf E}({\bf r'})
\end{equation}
where $ \varepsilon({\bf r},{\bf r'}) $ is the dielectric permittivity tensor. 
Equivalently to eq.($\ref{nonlocal}$) one can express the nonlocal 
relationship in terms of the polarization fields \cite{korn4}. 

While the theory of nonlocal electrostatics remains firmly embedded within 
the well-understood framework of Maxwell's theory, it introduces a new
characteristic length scale absent in local electrostatics: the 
correlation length $\lambda$ of the polarization correlations between the 
solvent molecules. This length sets the relevant scale for the deviation 
of the dielectric properties of the solvent from its bulk value. Thus,
nonlocal electrostatics is a serious candidate for a more realistic
description of solvent properties, provided it is also computationally 
tractable. It is here where the difficulties arise, however.
The theory of nonlocal electrostatics, discussed in detail below, is 
technically considerably more demanding than local electrostatics,
since it is usually formulated as a system of coupled integro-differential 
equations.
Consequently, it has so far only been applied to idealized situations, 
and even then typically after introducing additional approximations in
order to obtain analytical results \cite{levadny,cherep}. 
For complex geometries, the solution of
the equations by numerical methods becomes a formidable task.  

Thus, a reformulation of the equations of nonlocal electrostatics is needed
in order to make the theory applicable to real-world problems. 
Here we present a scheme which allows to rewrite the theory in terms of a 
system of partial differential equations for the local fields which 
consequently makes it amenable to standard methods of numerical analysis. 
This derivation relies on two assumptions which are typically valid for
e.g. the discussion of solvation problems for biomolecules: 
(i) the linearity of the relationship between
dielectric response and the electric field, and (ii) the representation of
the dielectric function in terms of Green functions of known differential
operators. We first derive the set of equations, and then illustrate how to 
solve them on (simple) examples.

%%%%%%%% FIG 1  %%%%%%%%%%%%%%%%%%%%%%%%%%%%%%%%%%%%%%%%%%%%%%%%%%%%%%%%%
\begin{figure}[t]
\includegraphics[height=5cm]{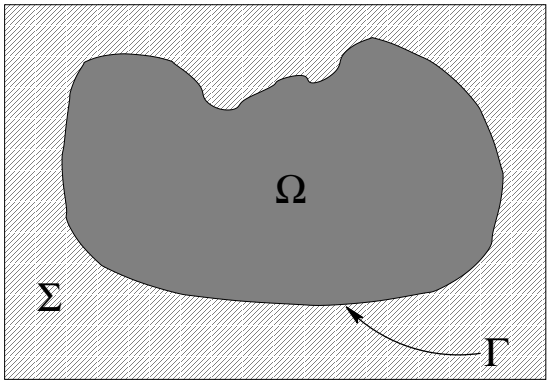}
\caption{A cavity with local dielectric constant ($\Omega$) immersed in a 
medium with spatially varying permittivity ($\Sigma$). The interface between
the two spaces is denoted by $\Gamma$.}
\label{FIG01}
\end{figure}
%%%%%%%% FIG 1  %%%%%%%%%%%%%%%%%%%%%%%%%%%%%%%%%%%%%%%%%%%%%%%%%%%%%%%%%

In the following we consider the situation given in Figure 1. 
A domain $ \Omega $ (which represents a molecule) is embedded in 
the solvent $\Sigma$ which takes up the whole space except $ \Omega$. 
The surface of the embedded domain is denoted by $\Gamma$. Within $\Omega$, 
the dielectric properties are assumed to be local, i.e., 
$\varepsilon \equiv \varepsilon_{\Omega} $, while in the solvent space 
$\Sigma$, eq.(\ref{nonlocal}) holds. On large scales, i.e. when
$|{\bf r} - {\bf r'}| \gg \lambda $, the dielectric response in the solvent 
is again local with the bulk dielectric constant $ \varepsilon_{\Sigma} $
(with, e.g., $ \varepsilon_{\Sigma} \approx 80$ for water).  
In this situation the standard equations of nonlocal electrostatics read as 
\begin{equation} \label{poissono}
\Delta \phi_{\Omega} 
=  - \frac{1}{\varepsilon_{\Omega}\varepsilon_{0}}\varrho\,\,\,, 
\,\,\,\,\,\,\,{\bf r}\,\, \epsilon\,\, \Omega 
\end{equation}
\begin{equation} \label{poissons}
\varepsilon_0\nabla\int_{\Sigma}d{\bf r'}\varepsilon({\bf r},{\bf r'})
\nabla'\phi_{\Sigma}({\bf r'})
 =  0\,,\,\,\,\,\,\,\,\,\,\,\,\,\,{\bf r},{\bf r'}\,\, \epsilon\,\, \Sigma 
\end{equation}
where $\varrho$ is the density of fixed charges which are assumed to lie 
confined within $\Omega\backslash\Gamma $, which is usually the case for
biomolecules in solution. This is no restriction on the
validity of our approach: the existence of surface charges on $\Gamma$
only slightly modifies the boundary conditions. 
Eqs.(\ref{poissono},\ref{poissons}) are the 
Poisson equations for the geometry of Fig.~1. 
The primed symbol $\nabla'$ denotes the differentiation with respect 
to ${\bf r'}$. 
Without surface charges on $\Gamma$, the boundary conditions 
to eqs.(\ref{poissono},\ref{poissons}) 
for the normal (n) and tangential (t) components of the electric 
and dielectric displacement field on the boundary $\Gamma$ are given by
\begin{equation} \label{boundary}
{\bf D}_{\Omega,n} = {\bf D}_{\Sigma,n}\,\,\,,\,\,\, 
{\bf E}_{\Omega,t} = {\bf E}_{\Sigma,t} 
\end{equation}
%\begin{eqnarray} \label{boundary}
%\left[\int_{\Sigma}d{\bf r'}\varepsilon({\bf r},{\bf r'})\nabla'\phi_{\Sigma}({\bf r'}) 
%- \varepsilon_{\Omega}\nabla\phi_{\Omega}({\bf r}) \right]\cdot 
%\widehat{\bf n}(\bf{r}) 
%& = & 0\,\,,
%{\bf r}\,\,\, \epsilon\,\, \Gamma \nonumber \\
%\\
%\left[\nabla\phi_{\Sigma} - \nabla\phi_{\Omega} \right]\times 
%\widehat{\bf n}(\bf{r}) & = & 0\,\,, {\bf r}\,\,\, 
%\epsilon\,\, \Gamma \nonumber \\
%\end{eqnarray}   
%where primed symboles $\nabla'$ denote the differentiation with respect to
%$ {\bf r'} $, and $ \widehat{\bf n}$ is the unit normal vector to $\Gamma$.
where, by virtue of eq.(\ref{poissons}), the boundary condition for $\bf{D}$ is
also nonlocal.

In order to step over from the integro-differential to a purely differential 
formulation we introduce, in addition to the potential field $ \phi $, another
potential field $ \psi $ within both compartments $\Omega$ and $\Sigma$.
Attempts similar in spirit, but differing in the implementation, have been
discussed before in the literature \cite{basil1}, again resulting in systems of
integro-differential equations.  
In $\Omega$ we define the relations between the potentials $\phi$ and $\psi$ 
and the physical fields as
\begin{equation} \label{inomega}
{\bf E}_{\Omega} \equiv -\nabla\phi_{\Omega}\,\,,
\,\,\,{\bf D}_{\Omega} \equiv -\nabla\psi 
\end{equation}
while in $\Sigma$ we have
\begin{equation}  \label{insigmae}
{\bf E}_{\Sigma} \equiv -\nabla\phi_{\Sigma_{\Omega}}
\end{equation}
while the dielectric displacement field in $\Sigma$ can be represented
in terms of a scalar and a vector field \cite{arfken},
\begin{equation} \label{insigmad}
{\bf D}_{\Sigma} \equiv 
-\varepsilon_0\int_{\Sigma}d{\bf r'} 
\varepsilon({\bf r},{\bf r'})\nabla'\phi_{\Sigma}
\equiv - \nabla \psi_{\Sigma} + \nabla\!\times\!{\bf\xi}_\Sigma\, .
\end{equation}
The scalar field $\psi$ thus serves as the potential of the rotation free part 
${\bf \tilde{D}}=-\nabla\psi_{\Sigma}$ of the dielectric displacement field ${\bf D}$.
In our setting $\phi_{\Sigma}$ is determined by $\psi_\Sigma$
alone, and neither $\psi_\Omega$ nor $\phi_\Omega$ are affected by 
${\bf\xi}_\Sigma$. Note that ${\bf\xi}_\Sigma$ can of
course be computed from e.g. eq. (\ref{insigmad}) as soon as $\phi_\Sigma$ is 
known. Since we here are interested only in the electrostatic potential 
$\phi$ and quantities derived from it, we will ignore ${\bf\xi}_\Sigma$ 
in the following.

With this definition, the differential equations and boundary conditions, 
eqs.(\ref{poissono} - \ref{insigmad}) can be brought into the form
\begin{equation} \label{poissonpsi}
\Delta\psi_{\Omega} = - \varrho\,\,\,,
\,\,\,\,{\bf r}\,\, \epsilon\,\, \Omega\,\,\,,\,\,\, \Delta\psi_{\Sigma} = 0 
\,\,,\,\,\,{\bf r}\,\, \epsilon\,\, \Sigma 
\end{equation}
and
\begin{equation} \label{psibdry} 
\nabla\psi_{\Sigma}|_n = \nabla\psi_{\Omega}|_n\,\,,\,\,
{\bf r}\,\, \epsilon\,\, \Gamma\,\,,
\nabla\phi_{\Sigma}|_t = \nabla\phi_{\Omega}|_t\,\,,\, 
{\bf r}\,\, \epsilon\,\, \Gamma
\end{equation}
In addition, there are now two equations relating the potential fields
$\phi$ and $\psi$, 
\begin{equation}  \label{psiphi}
\varepsilon_0\varepsilon_{\Omega}\phi_{\Omega} = \psi_{\Omega}
\,\,,\,\,\,\,{\bf r}\,\, \epsilon\,\, \Omega 
\end{equation}
and eq.(\ref{insigmad}), to be fulfilled in $\Sigma$.

So far, the introduction of $\psi$ has made it possible to rewrite the
boundary conditions in a completely local way.  
The nonlocal problem now only is with eq.(\ref{insigmad}) 
in which the nonlocal integral kernel still remains.
The latter expression can be simplified further under additional 
assumptions on the explicit form of $\varepsilon({\bf r},{\bf r'})$.
We now assume that it can be written in the form
\begin{equation} \label{epsilon}
\varepsilon({\bf r},{\bf r'}) = \varepsilon_{\ell}
\delta({\bf r} - {\bf r'})
+ \tilde{\varepsilon}{\cal G}({\bf r},{\bf r'}) 
\end{equation}
where 
$\tilde{\varepsilon} = (\varepsilon_{\Sigma} - \varepsilon_{\ell})/\lambda^2 
> 0$, and the Green function $\cal G$ solves 
\begin{equation} \label{green}
\cal{L}\, G({\bf r},{\bf r'}) = -\delta({\bf r} - {\bf r'})
\end{equation} 
for a differential operator $\cal{L}$ with constant coefficients. 
Note that $\varepsilon_{\ell}$ refers
to the value of the dielectric function on smallest scales (i.e., 
$ {\bf r} \rightarrow {\bf r'} $) and can be related to the frequency 
spectrum of the dielectric function \cite{korn4}. While this construction
clearly restricts the theory of nonlocal electrostatics to a certain class 
of dielectric functions, this restriction is not problematic as it applies to
most situations of physical interest.

Under these assumptions the application
of the differential operator $\cal L$ to eq.(\ref{insigmad}) yields
\begin{equation} \label{insigmadl}
\varepsilon_0\left[\varepsilon_{\ell}{\cal L} - \tilde{\varepsilon}\right]
\nabla \phi_{\Sigma} = -{\cal L}\nabla \psi_{\Sigma}\, .
\end{equation} 
As the fields $\phi$ and $\psi$ are determined only up to an arbitrary 
constant,
we can drop the gradients on both sides of eq.(\ref{insigmadl}) and are then 
left with the expression
\begin{equation} \label{phipsieq}
\varepsilon_0\left[\varepsilon_{\ell}{\cal L} - \tilde{\varepsilon}\right]
\phi_{\Sigma} = -{\cal L}\psi_{\Sigma}\, .
\end{equation}
Eqs.(\ref{poissonpsi} - \ref{psiphi},\ref{phipsieq}) 
constitute the novel formulation of nonlocal electrostatics based entirely 
on partial differential equations. 

In order to apply the equations to specific physical situations, the nonlocal 
dielectric function and thus the Green
function of the differential operator $\cal L$ need to be determined. 
To keep the computations simple -- since here we focus only on the
basic conceptual features of our approach --  we further assume  
that the dielectric function is isotropic, i.e., 
${\cal G}({\bf r},{\bf r'}) \approx {\cal G}({\bf r} - {\bf r'})$, 
which is exact far from any boundary. The use of more general
expressions is clearly permitted in our theory and is in fact needed for
the treatment of realistic situations \cite{ritschel,attard,korn5}. Within
our approach, they lead to more complex Green functions and corresponding
differential operators, and will be discussed in a detailed study 
later. %\cite{anhi2}. 

A standard model for an isotropic nonlocal dielectric function is
given by the so-called Fourier-Lorentzian model with a Yukawa-type kernel
in real space. The corresponding Green function reads
\begin{equation} \label{yukawa}
{\cal G}({\bf r} - {\bf r'}) = \frac{1}{4\pi}
\frac{e^{- {\frac{|{\bf r} - {\bf r'}|}{\lambda}}}}
{|{\bf r} - {\bf r'}|}
\end{equation}
with the differential operator ${\cal L} $ being given by $ {\cal L} \equiv
\Delta - 1/\lambda^2 $.
With this choice eq.(\ref{phipsieq}) reads
\begin{equation} \label{hs4}
\varepsilon_0\left[\varepsilon_{\ell}\lambda^2 \Delta - \varepsilon_{\Sigma}
\right]\phi_{\Sigma} = \psi_{\Sigma}
\end{equation}
Note that due to $\Delta \psi_{\Sigma} = 0 $,
no differential operator appears on the rhs of eq.(\ref{hs4}).

This result is interesting for two reasons. First, it
illustrates that in the limit $\lambda \rightarrow 0 $, i.e. on length scales
large compared to the scale of the orientational correlations,
the local limit is recovered.
Second, for the differential operator chosen, the form of the equation
is apparently that of a Debye-H\"uckel equation in which the role of
the Debye-H\"uckel screening length is played by the combination
$\lambda(\varepsilon_{\ell}/\varepsilon_{\Sigma})^{\, 1/2}$,
and the potential $\psi$ plays the role of the density of
mobile charges. We can thus interpret $\psi$ as a density of polarization
charges in the bulk whose gradient gives rise to the rotation free part of
the displacement field ${\bf D}$.

We now turn to illustrate how our formulation of 
nonlocal electrostatics can be put to use. First, we consider the simplest 
case of a charge $q$ placed at the center of a spherical shell of radius $a$.
Inside the shell, we assume $ \varepsilon_{\Omega} = 1 $. This system serves
as a model for ion solvation \cite{korn4}. 
The equations for $\psi$ and $\phi$ can now be solved as follows.
In the nonlocal case, the role of the Poisson equation for $\phi$ is
taken over by the equation for $\psi$ according to
eqs.(\ref{poissonpsi},\ref{psibdry}).
The $\psi$-potential inside the shell is given by $\psi_{\Omega} = \frac{q}{4\pi r} $,
with the same form given outside, $\psi_{\Sigma} = \frac{q'}{4\pi r}$. 
The tangential boundary condition is trivially fulfilled, while the 
normal boundary condition at $ r= a $ leads to $q=q'$.
Due to the radial symmetry of the problem, eq. (\ref{hs4}) reads 
\begin{equation} \label{phipsiion}       
\varepsilon_0\left[\varepsilon_{\ell}\lambda^2\left(\frac{1}{r^2}\frac{\partial}{\partial_r}\left(r^2\frac{\partial}{\partial_r}\right)\right) - \varepsilon_{\Sigma}\right]
\phi_{\Sigma} = \frac{q}{4\pi r}
\end{equation}
which is solved by
\begin{equation} \label{phiion}
        \phi_{\Sigma}(r) = \frac{1}{4\pi\varepsilon_\Sigma\varepsilon_0}\frac{q}{r}(1 + A\exp(-\gamma r))
\end{equation}
where the coefficients $A$ and $\gamma$ follow from coefficient
matching and the continuity of $\phi$ at the boundary, 
$ \phi_{\Sigma}(a) = \phi_{\Omega}(a)$
\begin{eqnarray}
A \equiv \frac{\varepsilon_\Sigma - \varepsilon_l}{\varepsilon_l}
\frac{\sinh{\nu}}{\nu}\,\,,\,
\nu \equiv\sqrt{\frac{\varepsilon_\Sigma}{\varepsilon_l}}\frac{a}{\lambda}
\,\,,\, \gamma \equiv\frac{\nu}{a} = 
\sqrt{\frac{\varepsilon_\Sigma}{\varepsilon_l}}\frac{1}{\lambda} \nonumber \\
& &
\end{eqnarray}
The electrostatic potential can be used to estimate the solvation energy of 
monovalent and divalent ions. From $\phi_\alpha(r)$ and $\psi_\alpha(r)$, 
with $\alpha\in \{\Omega, \Sigma\}$, we can easily compute
the free energy of solvation for this setting as the difference of 
the electrostatic energies in water and vacuum (where the local computation 
of course remains valid) from $ \Delta G=\frac{1}{2}\int\{\rho\phi - 
\rho\phi_{\text{vac}}\}dr$,
where the integrals are split into integrals over $\Omega$ and $\Sigma$.

The result of this calculation is shown in Figure 2, where we have compared 
our results to a corresponding local computation (the Born-model).
The correlation length $\lambda $ serves as an adjustment
parameter; it is the only one in the theory. Like 
the ion radii, it can also be obtained from microscopic simulations.
Our result compares favourably to the experimental
data taken from \cite{exp}. The value for $\lambda$ was taken to be $24.13$\AA; for
the ion radii we chose the values according to {\AA}qvist
\cite{AAqvist}. We also tested the set of Shannon radii \cite{Shannon1, Shannon2} 
without significant differences on our results. A detailed discussion 
of the choice of ion radii within nonlocal electrostatics can be 
found in \cite{arxiv}.
%%%%%%%% Fig 2 %%%%%%%%%%%%%%%%%%%%%%%%%%%%%%%%%%%%%%%%%%%%%%%%%%%%%%
\begin{figure}[t]
\includegraphics[height=7cm]{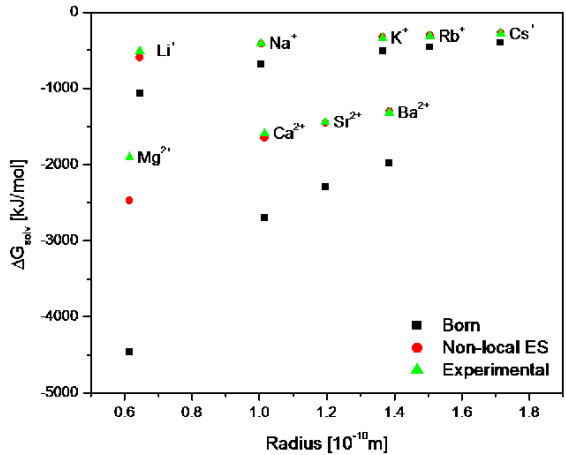}
\caption{Solvation free energy of ions.}
\label{FIG02}
\end{figure}
%%%%%%%% FIG 2  %%%%%%%%%%%%%%%%%%%%%%%%%%%%%%%%%%%%%%%%%%%%%%%%%%%%%%%%

The simple radially symmetric problem of ion solvation is, of course, not  
representative for the general character of the solutions of eq.(\ref{hs4}),
as the tangential boundary condition is trivially fulfilled in this case. 
In order to elucidate the effect of this boundary condition, we consider 
the potential generated by a charge 
$q$ placed at a distance $d$ from a planar dielectric phase boundary, 
as sketched in Figure 3.
%%%%%%%% FIG 3  %%%%%%%%%%%%%%%%%%%%%%%%%%%%%%%%%%%%%%%%%%%%%%%%%%%%%%%%%
\begin{figure}[t]
\includegraphics[height=4cm]{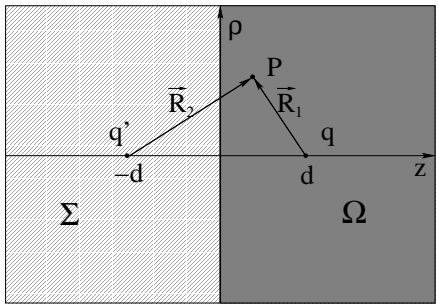}
\caption{Geometry of the problem of a charge $q$ placed at a distance 
$d$ from a dielectric boundary. For a discussion, see the text.}
\label{FIG03}
\end{figure}
%%%%%%%% FIG 3  %%%%%%%%%%%%%%%%%%%%%%%%%%%%%%%%%%%%%%%%%%%%%%%%%%%%%%%%%
Within local electrostatics, this problem can be solved by the method of 
image charges. The symmetry of the situation allows the use of cylindrical 
coordinates. At a point $P$ inside $\Omega$, the local dielectric medium, the 
electrostatic potential $\phi$ is given by 
%\begin{equation}  \label{hs1}
$\phi_{\Omega}(\rho,z) = - \frac{1}{4\pi\varepsilon_0\varepsilon_{\Omega}}
\left(\frac{q}{R_1} + \frac{q'}{R_2}\right)$
%\end{equation} 
where $R_1 = \sqrt{\rho^2 + (d-z)^2} $ and $R_2 = \sqrt{\rho^2 + (d+z)^2} $
are the lengths of the vectors pointing from the charge $q$ located at $z=d$, 
and the image charge $q'$ at $z = - d$ to the point $P$. 
Inside the charge-free halfspace $\Sigma $ the potential is given by 
%\begin{equation} \label{hs2}
$ \phi_{\Sigma} = 
\frac{1}{4\pi\varepsilon_0\varepsilon_{\Sigma}}\cdot \frac{q''}{R_1}$ 
%\end{equation}
with the image charge $q''$. 
The application of the boundary conditions for the normal and tangential
components then yields the relations between the image charges
%\begin{equation} \label{hs3}
$q + q' = 
\frac{\varepsilon_{\Omega}}{\varepsilon_{\Sigma}}q''\,\,\,,\,\,\,\ q - q' = q''$
%\end{equation}
so that the potential is determined explicitly in both halfspaces.
Inside $\Omega$, e.g., one has
\begin{equation} \label{phiomega2d}
\phi_{\Omega}(\varrho,z) = \frac{q}{4\pi\varepsilon_0\varepsilon_{\Omega}}
\left(
\frac{1}{R_1} + 
\frac{\varepsilon_{\Omega} - \varepsilon_{\Sigma}}{\varepsilon_{\Omega} + 
\varepsilon_{\Sigma}}\frac{1}{R_2}
\right)
\end{equation}
In the nonlocal case, the eqs. for $\psi$ have the same form as those
for $\phi$ in the local case, so we assume the solutions to be similar.
Under this assumption, the normal boundary condition remains unchanged 
while the tangential boundary condition leads to 
\begin{equation}
\partial_{\varrho} \phi_{\Sigma}= \frac{1}{4\pi\varepsilon_0\varepsilon_{\Omega}}
\frac{(q + q')}{[\varrho^2 + d^2]^{\, 3/2}}\varrho 
\end{equation}
at $ z = 0 $. This equation can be readily integrated along $\varrho$ and
determines the potential at the dielectric boundary. 
For $\varrho, z \rightarrow\infty $, on the other hand, $\phi_{\Sigma}$ must
vanish. These two conditions then allow to compute the potential 
$\phi_{\Sigma}$ from eq.(\ref{hs4}) within $\Sigma$. Since this 
requires a numerical computation,
we leave it for a future publication where we discuss the numerical 
treatment of our equations. % \cite{anhi2}.
Here, we only give the lowest order effect nonlocality within
$\Sigma$ has on $\Omega$. 
For $ \varrho \ll \sqrt{2}d$, the electrostatic potential inside $\Omega$ 
has the same form as in the local theory, but with a `renormalized' bulk 
value of the dielectric function $ \varepsilon_{\Sigma} = 
\varepsilon_{\Sigma,loc} - 2 \varepsilon_{\ell}(\lambda^2/d^2) $.
Transverse variations of the permittivity in the 
nonlocal medium thus induce a change of the local permittivity 
proportional to $(\lambda/d)^2$ in the vicinity of the boundary, an
effect which vanishes for $ \lambda \rightarrow 0 $, and for deeply
buried charges, $ d \rightarrow \infty$.

As a final remark on the applications of the nonlocal theory of 
electrostatics we mention the case in which mobile charges are present 
in the medium surrounding the cavity. Within the linear mean-field theory 
of local electrostatics, they can be described by Poisson-Boltzmann theory. 
This stays true within the nonlocal theory presented here. 
The Boltzmann distribution of the charges simply modifies the rhs of 
eq.(\ref{poissonpsi}). The nonlocal approach can then e.g. be used to quantify  
recent experimental results of AFM measurements on force-deflection 
curves at charged mica substrates in water and solutions of monovalent 
ions \cite{Teschke01}. 
Within a simplified treatment of the dielectric function of water, 
the orientational effects governed by $\lambda $ are found to be 
on the order of 10 nm \cite{Blossey03}. We expect that a more realistic 
structural model for water, which will become computationally tractable 
due to our approach, will lead to a much improved description of water 
orientation near charged surfaces, at least in cases where the assumption of 
mean-field behaviour is justified. 
  
To conclude, we have presented a novel formulation of nonlocal electrostatics,
which includes the effects of spatial dispersion in the dielectric permittivity
on surfaces embedded in a solvent, by reformulating it in terms of a 
two-potential model. While the resulting equations still need to be solved 
numerically even for simple geometries, this task can now be performed by
standard methods developed for partial differential equations. 
Due to the generality of eq.(\ref{phipsieq}), dielectric functions of 
greater complexity than the simple radially symmetric choice used here 
for illustrative purposes can be treated, provided they
can be related to known Green functions. Work in this direction
is under way.

{\bf Acknowledgement.} We thank the DFG for support under its research cluster 
``Informatics methods for the analysis and interpretation of large genomic datasets'',
grant LE952/2-3.

\end{document}